\documentclass[preprint,showpacs,preprintnumbers,amsmath,amssymb]{revtex4}
\usepackage{citesort}
\usepackage{epsfig}
\usepackage{graphicx}
\usepackage{dcolumn}
\usepackage{bm}
\usepackage{amsmath}
\newcommand{\be}{\begin{equation}}
\newcommand{\ee}{\end{equation}}

\begin{document}

\preprint{APS/123-QED}

\title{A Time-Orbiting Potential Trap\\for Bose-Einstein Condensate Interferometry}
\author{J.M. Reeves, O. Garcia, B. Deissler, K.L. Baranowski, K.J. Hughes, and C.A. Sackett}%
\affiliation{Physics Department, University of Virginia,
Charlottesville, VA 22904}
\email{jmr5p@virginia.edu}
\date{\today}

\begin{abstract}
We describe a novel atom trap for Bose-Einstein condensates of
$^{87}$Rb to be used in atom interferometry experiments. The trap is
based on a time-orbiting potential waveguide. It supports the atoms
against gravity while providing weak confinement to minimize
interaction effects. We observe harmonic oscillation frequencies
($\omega_x$,$\omega_y$,$\omega_z$) as low as 2$\pi\times
(6.0,1.2,3.3)$~Hz. Up to $2\times 10^4$ condensate atoms have been
loaded into the trap, at estimated temperatures as low as 850~pK. We
anticipate that interferometer measurement times of 1 s or more
should be achievable in this device.
\end{abstract}

\pacs{39.20.+q, 03.75.-b, 03.75.Be}

\maketitle

Atom interferometry with Bose-Einstein condensates has drawn a
considerable amount of interest due to the potential for
high-precision measurements \cite{Bongs04}.  The fundamental limit on the
sensitivity of an atom-based Sagnac interferometer, for example, exceeds a
photon-based interferometer by a factor of 10$^{11}$, for
middle-weight atoms and optical-wavelength light. Sensor
applications are also more numerous for an atom interferometer,
given that atoms are affected by electric and magnetic fields while
photons are not. This heightened sensitivity also amplifies the
effects of environmental noise, imposing practical limits on the
sensitivity that can be obtained.  Nonetheless, the
best gyroscope on record is an atom interferometer \cite{Gustavson00}.

Using a Bose-Einstein condensate for interferometry 
is appealing because the intrinsic
advantages of thermal atoms are considerably increased.  Slow velocities and
long coherence lengths allow condensates to exhibit greater
sensitivity per atom than a thermal cloud with similar number of
atoms.  Though producing and working with condensates remains
challenging, several groups have demonstrated Mach-Zehnder or
Michelson interferometers using condensates
\cite{Andrews97,Hagley99,Torii00,Bongs01,Gupta02,Shin04,Wang05,Saba05}.
However, all these experiments have been limited to measurement times of
roughly ten milliseconds or less. In this paper we discuss a novel
atom waveguide that we expect will permit significantly longer
measurement times.

In the design of a condensate interferometer, one must decide how
the atoms will be transported through the device.  The simplest
method is to orient the axis of the device vertically, allowing the
atoms to fall freely under the influence of gravity
\cite{Andrews97,Torii00,Gupta02}. While this technique introduces no
additional fields or dephasing effects, the measurement time is
limited by the speed at which the condensate falls.  However, BEC
experiments generically suffer from low production rates. This reduces the
signal-to-noise ratio, since the statistical fluctuations in phase
scale as $N^{-\frac{1}{2}}$ for number of atoms $N$. 
While thermal atomic beam experiments can
produce 10$^9$ atoms/s, condensate production rates are more
typically 10$^5$ atoms/s. In order to make up for these low numbers,
long interaction times will be required so that the overall phase is
increased.  This makes interferometers based on falling atoms
unattractive, though some of the difficulties might be circumvented using
either a fountain geometry \cite{Wynands05} or a magnetic
levitation approach \cite{Weber03}.

The alternative possibility is to use trapped atoms.  
Condensate interferometers using atoms confined by either magnetic
\cite{Hagley99,Wang05,Shin05} or optical
\cite{Bongs01,Shin04,Saba05} fields have been demonstrated.
Measurement times in these devices have been limited for a variety
of reasons, but a common concern is the effect of interatomic
interactions, which can introduce phase noise and cause spatial
distortions in the cloud \cite{Shin04,Saba05,Olshanii05}.
Confinement also imposes severe geometrical constraints due to the
need to avoid uncontrolled motional excitations \cite{Shin05,Wu05}.  To avoid
these problems, one wants a trap
capable of holding the atoms against gravity but otherwise as
weakly confining as possible.  Weak three-dimensional 
confinement has previously been observed \cite{Leanhardt03b}, but
in this paper we present a novel weakly confining
waveguide that is particularly well-suited
for the demands of atom interferometry.

The waveguide is illustrated in Fig.~1.  It is based on a
four-wire linear quadrupole and uses the time-orbiting potential
(TOP) technique \cite{Petrich95,Arnold02,Gupta05}.
Four current-carrying rods provide a linear
quadrupole field, with the zero line at the center. A rotating bias
field pushes the zero away from the atoms to prevent Majorana
losses.  We preferred the TOP to other options because of its
noise-reduction effects. Our bias field rotates at about 10 kHz, and
the atomic spins follow adiabatically. Because of this, any slowly
varying magnetic fields or other environmental noise coupling to the
spins will tend to average out. The atoms do
become sensitive to noise near 10 kHz, but we have found
that most of the magnetic noise in our lab has frequencies well
below this.

Conventional TOP traps are not especially weak.  The obvious way to
reduce the confinement strength is to reduce the magnetic field amplitude, but
we cannot lower the force below what is required to counter
gravity. The solution we have found is to oscillate not only the
bias field, but also the quadrupole field. With the appropriate
choice of phase, this causes the field zero to oscillate back and
forth above the atoms.  The atoms are  constantly attracted to the
overhead zero, and we can weaken the confinement further than would
otherwise be possible.

To understand this mechanism, suppose
the oscillating quadrupole field is
\begin{equation}
\mathbf{B}_Q = B_{Q}^{'}(x\mathbf{\hat{x}}-z\mathbf{\hat{z}})\cos\Omega t
\end{equation}
where $\mathbf{\hat{x}}$ and $\mathbf{\hat{z}}$ are the transverse
directions and $\mathbf{\hat{y}}$ is along the axis of the
waveguide. The $\mathbf{\hat{z}}$ direction is vertical. The bias
field is
\begin{equation} \mathbf{B}_0 =
B_0(\mathbf{\hat{x}}\sin\Omega t + \mathbf{\hat{z}}\cos\Omega t)
\end{equation}
with rotation frequency $\Omega = 11.9$ kHz. The resulting
time-averaged field magnitude, to second order in the coordinates, is
\begin{equation} \label{simpleavg}
\langle|\mathbf{B}|\rangle = B_0 - \frac{1}{2}B_{Q}^{'}z +
\frac{B_{Q}^{'2}}{16B_0}(3x^2 + z^2) ,
\end{equation}
providing a total potential energy
\begin{equation} U = \mu B + mgz = \mu B_0 -\frac{1}{2}\mu B_{Q}^{'}z + mgz
+ \frac{1}{2}m(\omega_x^2x^2 + \omega_z^2z^2) 
\end{equation}
for atomic mass $m$, gravitational acceleration $g$, and magnetic
moment $\mu$.
We set the
gradient $B_Q^{'}=2mg/\mu$ to support the atoms against gravity and
obtain trap frequencies
\begin{equation}
\omega_x = \left(\frac{3mg^2}{2\mu
B_0}\right)^{\frac{1}{2}}\quad \mbox{and} \quad ~ \omega_z =
\frac{\omega_x}{\sqrt3} .
\end{equation}
For a 10 G bias field,
this gives trap frequencies $\omega_x=2\pi\times7$ Hz and
$\omega_z=2\pi\times4$ Hz for Rb atoms in a state with maximum
$\mu$.

The waveguide is made of machined copper rods held inside a vacuum
chamber, as shown in Fig.~1.  Each of the four rods is a
coaxial pair.  As Fig.~2 illustrates, the four outer conductors
are connected in one circuit that provides the quadrupole field
while the inner conductors form two circuits used to generate the
two components of the bias field.
The end loops shown on the
quadrupole circuit help minimize the axial quadrupole field.

The leads of the circuits do have an appreciable
effect on the trap potential.
For instance, any residual axial component
causes the quadrupole field to become
\begin{equation} \label{axquad}
\mathbf{B}_Q = (ax\mathbf{\hat{x}}
+ cy\mathbf{\hat{y}}
- bz\mathbf{\hat{z}}
)\cos\Omega t
\end{equation}
with $c = b - a$.
A more accurate description
of the potential requires the inclusion of many other first- and
second-order terms in the magnetic fields.  Categorizing all
such terms is prohibitive,
but we have found that in the relevant parameter range,
the total average field magnitude is well approximated by
\begin{equation} \label{fullfield}
\langle|\mathbf{B}|\rangle = B_0 -
\frac{1}{2}bz
+ \left(\frac{3}{16B_0}a^2+\alpha B_0\right)x^2
+ \left(\frac{1}{4B_0}c^2 + \gamma B_0\right)y^2
+ \left(\frac{1}{16B_0}b^2+ \beta B_0\right)z^2
\end{equation}
which yields frequencies
\begin{equation}
\begin{split} \label{eq-freqs}
\omega_x &= \left[\frac{2\mu}{m}\left(\frac{3}{16B_0}a^2 + \alpha B_0\right)\right]^{\frac{1}{2}} \\
\omega_y &= \left[\frac{2\mu}{m}\left(\frac{1}{4B_0}c^2+\gamma B_0\right)\right]^{\frac{1}{2}} \\
\omega_z &= \left[\frac{2\mu}{m}\left(\frac{1}{16B_0}b^2 + \beta B_0\right)\right]^{\frac{1}{2}} .
\end{split}
\end{equation}
Here $a$, $b$ and $c$ are from Eq.~(\ref{axquad}) and $\alpha$,
$\beta$, and $\gamma$ come from variations in the bias fields. We
obtained this form by modeling the total field using the
Biot-Savart law and the mechanical design of the leads.  The model
predicts $B_0/I_0 = 0.40$ G/A, $a/I_Q = -0.83$~G/(A cm), $b/I_Q =
-0.86$~G/(A cm), $\alpha = -0.11$~cm$^{-2}$, $\beta =
-0.061$~cm$^{-2}$, and $\gamma = 0.019$~cm$^{-2}$, where $I_0$ is
the bias current amplitude and the $I_Q$ is the quadrupole current
amplitude. These values were coarsely verified using a gaussmeter,
yielding $B_0/I_0 \approx$ 0.4~G/A and 
$a/I_Q \approx b/I_Q \approx 0.8$~G/(A~cm).

The three trap circuits have similar impedances, presenting a
10~m$\Omega$ resistive and 0.3~$\mu$H inductive load.  The circuits
are driven with an actively stabilized commercial audio amplifier,
using transformers to match the amplifier's output impedance. The
details of this drive circuit will be presented elsewhere.  The trap
is mounted on several copper blocks that deliver the current and
remove heat. The measured thermal coefficient of the trap structure
is 2~W/K.  The quadrupole field requires a current of 38 A to cancel
gravity and a bias field of 20 G requires $I_0 = 50$ A in both bias
circuits, yielding a total temperature rise of about 16 K.

Our BEC system is based on the scheme described by Lewandowski {\it
et al.} \cite{Lewandowski03}.  We have a single MOT, separated from
an ultra-high vacuum science cell by a tube 30 cm long with diameter 1
cm. Our MOT contains 2$\times$10$^9$ $^{87}$Rb atoms at roughly
200~$\mu$K. We optically pump them into the $F = 2, m = 2$ ground
state for magnetic trapping.  We transfer the atoms to a spherical
quadrupole trap, obtaining about 1.5$\times$10$^9$ atoms at
900~$\mu$K with an axial field gradient of 387~G/cm.  The atoms are
transported to the science cell by a programmable motor, which moves
the electromagnet coils at $v=0.8$~m/s.

Once in place within the waveguide structure, we
evaporatively cool the cloud.  The atoms are initially too hot for
our TOP trap, so we start evaporating in the quadrupole trap. We
evaporate on the spin state transitions within the F=2 ground state
manifold.  Once the cloud cools below 200~$\mu$K, we turn on the
waveguide bias field and continue evaporating.  The static spherical
quadrupole field remains on to provide tight confinement.  We
achieve condensation with about 2$\times10^4$ atoms at a temperature
of 50~nK, using a 3.69~G bias field.  From the evaporative cooling, we
obtained a more accurate calibration of the bias field as 0.440 G/A
times the current amplitude $I_0$.

Our final atom number is somewhat lower than typical. We believe this
is because transferring the atoms from the quadrupole trap to the
TOP trap is inefficient and we are exploring ways to improve this.
Once the condensate is made the waveguide quadrupole field is ramped
on and the spherical quadrupole field ramped off.
The centers of the main trap and the waveguide do not exactly
coincide, so the fields are ramped with a 
7 s time period to enable the atoms to
move adiabatically to the new local B field minimum. We do not
observe any losses in the transfer. Figure~3 shows snapshots of the
cloud as the atoms are being transferred.

We measured the trap frequencies of the waveguide
by observing either center-of-mass (for the $x$ and $z$ directions) or
breathing mode (for the $y$ direction) oscillations in the cloud.
We perturbed the cloud by
introducing a sudden change in the confining field and then recorded
the subsequent behavior.  These tests were done on a noncondensed
cloud at temperatures of about 1~$\mu$K.
From the periods we determined the trap frequencies as a function of the
applied currents.

We measured the frequencies over a range of bias fields from 3 to
16~G. From this data, we solved for the trap parameters in our model
Eq.~(\ref{eq-freqs}). Using a multivariable minimization, we found
$|a|/I_Q = 0.734$~G/(A cm), $|b|/I_Q = 0.709$~G/(A cm), $\alpha =
0.17$~cm$^{-2}$, $\beta = 0.05$~cm$^{-2}$, and $\gamma =
0.02$~cm$^{-2}$.  The quadratic coefficients are rather different
from our model predictions, though the order of magnitude is
correct.  Using the empirical coefficients, Eq.~(\ref{eq-freqs})
reproduces the measured frequencies to an accuracy of about 0.1~Hz
over the range of bias fields tested.

The weakest confinement we observed, at $B_0 = 20.5$~G, had
$\omega_x = 2\pi\times6.0$~Hz, $\omega_y = 2\pi\times1.2$~Hz and
$\omega_z = 2\pi\times3.3$~Hz. By adiabatically expanding a small
condensate into such a weak trap, we were able to obtain very low
temperatures.  Figure 3(g) shows an image of 1.6$\times10^3$ atoms
in the trap with this bias field. We estimate the temperature of
this cloud to be 850~pK.  Although lower temperatures have been 
observed in Na \cite{Leanhardt03b}, this
is the lowest temperature achieved for Rb atoms of which we are aware.

With the successful demonstration of our trap, we are now preparing
to explore condensate interferometry.  We plan to conduct
experiments similar to those of Wang {\it et al.}, \cite{Wang05},
using a Bragg laser pulse to split and recombine condensate wave
packets.  The weak confinement of our guide should greatly reduce
the limiting effects of interactions. For instance, the phase
distortions discussed in \cite{Olshanii05} should have negligible
effect for condensate numbers below about $1.5\times 10^4$, and
phase diffusion effects during the wavepacket propagation should not
become important for interaction times less than about
1~s \cite{Javanainen97}.  Using the current apparatus, we plan
to study these and other limiting effects.  With suitable modifications,
our waveguide could be used to precisely measure electric polarizability,
gravitational forces, rotations, and other phenomena \cite{Berman94}.
We are hopeful that the trap design presented here
will help condensate interferometry realize this potential.

This work was supported by the US Office of Naval Research, the
National Science Foundation, the Research Corporation, and the
Alfred P. Sloan Foundation.

\pagebreak


\begin{thebibliography}{23}
\expandafter\ifx\csname natexlab\endcsname\relax\def\natexlab#1{#1}\fi
\expandafter\ifx\csname bibnamefont\endcsname\relax
  \def\bibnamefont#1{#1}\fi
\expandafter\ifx\csname bibfnamefont\endcsname\relax
  \def\bibfnamefont#1{#1}\fi
\expandafter\ifx\csname citenamefont\endcsname\relax
  \def\citenamefont#1{#1}\fi
\expandafter\ifx\csname url\endcsname\relax
  \def\url#1{\texttt{#1}}\fi
\expandafter\ifx\csname urlprefix\endcsname\relax\def\urlprefix{URL }\fi
\providecommand{\bibinfo}[2]{#2}
\providecommand{\eprint}[2][]{\url{#2}}

\bibitem[{\citenamefont{Bongs and Sengstock}(2004)}]{Bongs04}
\bibinfo{author}{\bibfnamefont{K.}~\bibnamefont{Bongs}} \bibnamefont{and}
  \bibinfo{author}{\bibfnamefont{K.}~\bibnamefont{Sengstock}},
  \bibinfo{journal}{Rep. Prog. Phys.} \textbf{\bibinfo{volume}{67}},
  \bibinfo{pages}{907} (\bibinfo{year}{2004}).

\bibitem[{\citenamefont{Gustavson et~al.}(2000)\citenamefont{Gustavson,
  Landragin, and Kasevich}}]{Gustavson00}
\bibinfo{author}{\bibfnamefont{T.~L.} \bibnamefont{Gustavson}},
  \bibinfo{author}{\bibfnamefont{A.}~\bibnamefont{Landragin}},
  \bibnamefont{and} \bibinfo{author}{\bibfnamefont{M.~A.}
  \bibnamefont{Kasevich}}, \bibinfo{journal}{Class. Quantum Grav.}
  \textbf{\bibinfo{volume}{17}}, \bibinfo{pages}{2385} (\bibinfo{year}{2000}).

\bibitem[{\citenamefont{Andrews et~al.}(1997)\citenamefont{Andrews, Townsend,
  Miesner, Durfee, Kurn, and Ketterle}}]{Andrews97}
\bibinfo{author}{\bibfnamefont{M.~R.} \bibnamefont{Andrews}},
  \bibinfo{author}{\bibfnamefont{C.~G.} \bibnamefont{Townsend}},
  \bibinfo{author}{\bibfnamefont{H.~J.} \bibnamefont{Miesner}},
  \bibinfo{author}{\bibfnamefont{D.~S.} \bibnamefont{Durfee}},
  \bibinfo{author}{\bibfnamefont{D.~M.} \bibnamefont{Kurn}}, \bibnamefont{and}
  \bibinfo{author}{\bibfnamefont{W.}~\bibnamefont{Ketterle}},
  \bibinfo{journal}{Science} \textbf{\bibinfo{volume}{275}},
  \bibinfo{pages}{637} (\bibinfo{year}{1997}).

\bibitem[{\citenamefont{Shin et~al.}(2004)\citenamefont{Shin, Saba, Pasquini,
  Ketterle, Pritchard, and Leanhardt}}]{Shin04}
\bibinfo{author}{\bibfnamefont{Y.}~\bibnamefont{Shin}},
  \bibinfo{author}{\bibfnamefont{M.}~\bibnamefont{Saba}},
  \bibinfo{author}{\bibfnamefont{T.~A.} \bibnamefont{Pasquini}},
  \bibinfo{author}{\bibfnamefont{W.}~\bibnamefont{Ketterle}},
  \bibinfo{author}{\bibfnamefont{D.~E.} \bibnamefont{Pritchard}},
  \bibnamefont{and} \bibinfo{author}{\bibfnamefont{A.~E.}
  \bibnamefont{Leanhardt}}, \bibinfo{journal}{Phys. Rev. Lett.}
  \textbf{\bibinfo{volume}{92}}, \bibinfo{pages}{050405}
  (\bibinfo{year}{2004}).

\bibitem[{\citenamefont{Hagley et~al.}(1999)\citenamefont{Hagley, Deng, Kozuma,
  Trippenbach, Band, Edwards, Doery, Julienne, Helmerson, Rolston
  et~al.}}]{Hagley99}
\bibinfo{author}{\bibfnamefont{E.~W.} \bibnamefont{Hagley}},
  \bibinfo{author}{\bibfnamefont{L.}~\bibnamefont{Deng}},
  \bibinfo{author}{\bibfnamefont{M.}~\bibnamefont{Kozuma}},
  \bibinfo{author}{\bibfnamefont{M.}~\bibnamefont{Trippenbach}},
  \bibinfo{author}{\bibfnamefont{Y.~B.} \bibnamefont{Band}},
  \bibinfo{author}{\bibfnamefont{M.}~\bibnamefont{Edwards}},
  \bibinfo{author}{\bibfnamefont{M.}~\bibnamefont{Doery}},
  \bibinfo{author}{\bibfnamefont{P.~S.} \bibnamefont{Julienne}},
  \bibinfo{author}{\bibfnamefont{K.}~\bibnamefont{Helmerson}},
  \bibinfo{author}{\bibfnamefont{S.~L.} \bibnamefont{Rolston}},
  \bibnamefont{et~al.}, \bibinfo{journal}{Phys. Rev. Lett.}
  \textbf{\bibinfo{volume}{83}}, \bibinfo{pages}{3112} (\bibinfo{year}{1999}).

\bibitem[{\citenamefont{Torii et~al.}(2000)\citenamefont{Torii, Suzuki, Kozuma,
  Sugiura, Kuga, Deng, and Hagley}}]{Torii00}
\bibinfo{author}{\bibfnamefont{Y.}~\bibnamefont{Torii}},
  \bibinfo{author}{\bibfnamefont{Y.}~\bibnamefont{Suzuki}},
  \bibinfo{author}{\bibfnamefont{M.}~\bibnamefont{Kozuma}},
  \bibinfo{author}{\bibfnamefont{T.}~\bibnamefont{Sugiura}},
  \bibinfo{author}{\bibfnamefont{T.}~\bibnamefont{Kuga}},
  \bibinfo{author}{\bibfnamefont{L.}~\bibnamefont{Deng}}, \bibnamefont{and}
  \bibinfo{author}{\bibfnamefont{E.~W.} \bibnamefont{Hagley}},
  \bibinfo{journal}{Phys. Rev. A} \textbf{\bibinfo{volume}{61}},
  \bibinfo{pages}{041602(R)} (\bibinfo{year}{2000}).

\bibitem[{\citenamefont{Bongs et~al.}(2001)\citenamefont{Bongs, Burger,
  Dettmer, Hellweg, Arlt, Ertmer, and Sengstock}}]{Bongs01}
\bibinfo{author}{\bibfnamefont{K.}~\bibnamefont{Bongs}},
  \bibinfo{author}{\bibfnamefont{S.}~\bibnamefont{Burger}},
  \bibinfo{author}{\bibfnamefont{S.}~\bibnamefont{Dettmer}},
  \bibinfo{author}{\bibfnamefont{D.}~\bibnamefont{Hellweg}},
  \bibinfo{author}{\bibfnamefont{J.}~\bibnamefont{Arlt}},
  \bibinfo{author}{\bibfnamefont{W.}~\bibnamefont{Ertmer}}, \bibnamefont{and}
  \bibinfo{author}{\bibfnamefont{K.}~\bibnamefont{Sengstock}},
  \bibinfo{journal}{Phys. Rev. A} \textbf{\bibinfo{volume}{63}},
  \bibinfo{pages}{031602(R)} (\bibinfo{year}{2001}).

\bibitem[{\citenamefont{Gupta et~al.}(2002)\citenamefont{Gupta, Dieckmann,
  Hadzibabic, and Pritchard}}]{Gupta02}
\bibinfo{author}{\bibfnamefont{S.}~\bibnamefont{Gupta}},
  \bibinfo{author}{\bibfnamefont{K.}~\bibnamefont{Dieckmann}},
  \bibinfo{author}{\bibfnamefont{Z.}~\bibnamefont{Hadzibabic}},
  \bibnamefont{and} \bibinfo{author}{\bibfnamefont{D.~E.}
  \bibnamefont{Pritchard}}, \bibinfo{journal}{Phys. Rev. Lett.}
  \textbf{\bibinfo{volume}{89}}, \bibinfo{pages}{140401}
  (\bibinfo{year}{2002}).

\bibitem[{\citenamefont{Wang et~al.}(2005)\citenamefont{Wang, Anderson, Bright,
  Cornell, Diot, Kishimoto, Prentiss, Saravanan, Segal, and Wu}}]{Wang05}
\bibinfo{author}{\bibfnamefont{Y.~J.} \bibnamefont{Wang}},
  \bibinfo{author}{\bibfnamefont{D.~Z.} \bibnamefont{Anderson}},
  \bibinfo{author}{\bibfnamefont{V.~M.} \bibnamefont{Bright}},
  \bibinfo{author}{\bibfnamefont{E.~A.} \bibnamefont{Cornell}},
  \bibinfo{author}{\bibfnamefont{Q.}~\bibnamefont{Diot}},
  \bibinfo{author}{\bibfnamefont{T.}~\bibnamefont{Kishimoto}},
  \bibinfo{author}{\bibfnamefont{M.}~\bibnamefont{Prentiss}},
  \bibinfo{author}{\bibfnamefont{R.~A.} \bibnamefont{Saravanan}},
  \bibinfo{author}{\bibfnamefont{S.~R.} \bibnamefont{Segal}}, \bibnamefont{and}
  \bibinfo{author}{\bibfnamefont{S.}~\bibnamefont{Wu}}, \bibinfo{journal}{Phys.
  Rev. Lett.} \textbf{\bibinfo{volume}{94}}, \bibinfo{pages}{090405}
  (\bibinfo{year}{2005}).

\bibitem[{\citenamefont{Saba et~al.}(2005)\citenamefont{Saba, Pasquini, Sanner,
  Shin, Ketterle, and Pritchard}}]{Saba05}
\bibinfo{author}{\bibfnamefont{M.}~\bibnamefont{Saba}},
  \bibinfo{author}{\bibfnamefont{T.~A.} \bibnamefont{Pasquini}},
  \bibinfo{author}{\bibfnamefont{C.}~\bibnamefont{Sanner}},
  \bibinfo{author}{\bibfnamefont{Y.}~\bibnamefont{Shin}},
  \bibinfo{author}{\bibfnamefont{W.}~\bibnamefont{Ketterle}}, \bibnamefont{and}
  \bibinfo{author}{\bibfnamefont{D.~E.} \bibnamefont{Pritchard}},
  \bibinfo{journal}{Science} \textbf{\bibinfo{volume}{307}},
  \bibinfo{pages}{1945} (\bibinfo{year}{2005}).

\bibitem[{\citenamefont{Wynands and Weyers}(2005)}]{Wynands05}
\bibinfo{author}{\bibfnamefont{R.}~\bibnamefont{Wynands}} \bibnamefont{and}
  \bibinfo{author}{\bibfnamefont{S.}~\bibnamefont{Weyers}},
  \bibinfo{journal}{Metrologia} \textbf{\bibinfo{volume}{42}},
  \bibinfo{pages}{S64} (\bibinfo{year}{2005}).

\bibitem[{\citenamefont{Weber et~al.}(2003)\citenamefont{Weber, Herbig, Mark,
  N{\"a}gerl, and Grimm}}]{Weber03}
\bibinfo{author}{\bibfnamefont{T.}~\bibnamefont{Weber}},
  \bibinfo{author}{\bibfnamefont{J.}~\bibnamefont{Herbig}},
  \bibinfo{author}{\bibfnamefont{M.}~\bibnamefont{Mark}},
  \bibinfo{author}{\bibfnamefont{H.-C.} \bibnamefont{N{\"a}gerl}},
  \bibnamefont{and} \bibinfo{author}{\bibfnamefont{R.}~\bibnamefont{Grimm}},
  \bibinfo{journal}{Science} \textbf{\bibinfo{volume}{299}},
  \bibinfo{pages}{232} (\bibinfo{year}{2003}).

\bibitem[{\citenamefont{Shin et~al.}(2005)\citenamefont{Shin, Sanner, Jo,
  Pasquini, Saba, Ketterle, Pritchard, Vengalattore, and Prentiss}}]{Shin05}
\bibinfo{author}{\bibfnamefont{Y.}~\bibnamefont{Shin}},
  \bibinfo{author}{\bibfnamefont{C.}~\bibnamefont{Sanner}},
  \bibinfo{author}{\bibfnamefont{G.-B.} \bibnamefont{Jo}},
  \bibinfo{author}{\bibfnamefont{T.~A.} \bibnamefont{Pasquini}},
  \bibinfo{author}{\bibfnamefont{M.}~\bibnamefont{Saba}},
  \bibinfo{author}{\bibfnamefont{W.}~\bibnamefont{Ketterle}}, 
  \bibinfo{author}{\bibfnamefont{D.~E.} \bibnamefont{Pritchard}},
  \bibinfo{author}{\bibfnamefont{M.} \bibnamefont{Vengalattore}},
  \bibnamefont{and}
  \bibinfo{author}{\bibfnamefont{M.} \bibnamefont{Prentiss}},
  \bibinfo{journal}{Phys. Rev. A} \textbf{\bibinfo{volume}{72}},
  \bibinfo{pages}{021604(R)} (\bibinfo{year}{2005}).

\bibitem[{\citenamefont{Olshanii and Dunjko}(2005)}]{Olshanii05}
\bibinfo{author}{\bibfnamefont{M.}~\bibnamefont{Olshanii}} \bibnamefont{and}
  \bibinfo{author}{\bibfnamefont{V.}~\bibnamefont{Dunjko}},
  \bibinfo{note}{eprint cond-mat/0505358}.

\bibitem[{\citenamefont{Wu et~al.}(2005)\citenamefont{Wu, Su, and
  Prentiss}}]{Wu05}
\bibinfo{author}{\bibfnamefont{S.}~\bibnamefont{Wu}},
  \bibinfo{author}{\bibfnamefont{E.~J.} \bibnamefont{Su}}, \bibnamefont{and}
  \bibinfo{author}{\bibfnamefont{M.}~\bibnamefont{Prentiss}},
  \bibinfo{journal}{Euro. Phys. J. D} \textbf{\bibinfo{volume}{35}},
  \bibinfo{pages}{111} (\bibinfo{year}{2005}).

\bibitem[{\citenamefont{Leanhardt
  et~al.}(2003{\natexlab{a}})\citenamefont{Leanhardt, Pasquini, Saba,
  Schirotzek, Shin, Kielpinski, Pritchard, and Ketterle}}]{Leanhardt03b}
\bibinfo{author}{\bibfnamefont{A.~E.} \bibnamefont{Leanhardt}},
  \bibinfo{author}{\bibfnamefont{T.~A.} \bibnamefont{Pasquini}},
  \bibinfo{author}{\bibfnamefont{M.}~\bibnamefont{Saba}},
  \bibinfo{author}{\bibfnamefont{A.}~\bibnamefont{Schirotzek}},
  \bibinfo{author}{\bibfnamefont{Y.}~\bibnamefont{Shin}},
  \bibinfo{author}{\bibfnamefont{D.}~\bibnamefont{Kielpinski}},
  \bibinfo{author}{\bibfnamefont{D.~E.} \bibnamefont{Pritchard}},
  \bibnamefont{and} \bibinfo{author}{\bibfnamefont{W.}~\bibnamefont{Ketterle}},
  \bibinfo{journal}{Science} \textbf{\bibinfo{volume}{301}},
  \bibinfo{pages}{1513} (\bibinfo{year}{2003}{\natexlab{a}}).

\bibitem[{\citenamefont{Petrich et~al.}(1995)\citenamefont{Petrich, Anderson,
  Ensher, and Cornell}}]{Petrich95}
\bibinfo{author}{\bibfnamefont{W.}~\bibnamefont{Petrich}},
  \bibinfo{author}{\bibfnamefont{M.~H.} \bibnamefont{Anderson}},
  \bibinfo{author}{\bibfnamefont{J.~R.} \bibnamefont{Ensher}},
  \bibnamefont{and} \bibinfo{author}{\bibfnamefont{E.~A.}
  \bibnamefont{Cornell}}, \bibinfo{journal}{Phys. Rev. Lett.}
  \textbf{\bibinfo{volume}{74}}, \bibinfo{pages}{3352} (\bibinfo{year}{1995}).

\bibitem[{\citenamefont{Arnold and Riis}(1999)}]{Arnold02}
\bibinfo{author}{\bibfnamefont{A.~S.} \bibnamefont{Arnold}} \bibnamefont{and}
  \bibinfo{author}{\bibfnamefont{E.}~\bibnamefont{Riis}}, \bibinfo{journal}{J.
  Mod. Optics} \textbf{\bibinfo{volume}{49}}, \bibinfo{pages}{5861}
  (\bibinfo{year}{1999}).

\bibitem[{\citenamefont{Gupta et~al.}(2005)\citenamefont{Gupta, Murch, Moore,
  Purdy, and Stamper-Kurn}}]{Gupta05}
\bibinfo{author}{\bibfnamefont{S.}~\bibnamefont{Gupta}},
  \bibinfo{author}{\bibfnamefont{K.~W.} \bibnamefont{Murch}},
  \bibinfo{author}{\bibfnamefont{K.~L.} \bibnamefont{Moore}},
  \bibinfo{author}{\bibfnamefont{T.~P.} \bibnamefont{Purdy}}, \bibnamefont{and}
  \bibinfo{author}{\bibfnamefont{D.~M.} \bibnamefont{Stamper-Kurn}},
  \bibinfo{note}{eprint cond-mat/0504749}.

\bibitem[{\citenamefont{Lewandowski et~al.}(2003)\citenamefont{Lewandowski,
  Harber, Whitaker, and Cornell}}]{Lewandowski03}
\bibinfo{author}{\bibfnamefont{H.~J.} \bibnamefont{Lewandowski}},
  \bibinfo{author}{\bibfnamefont{D.~M.} \bibnamefont{Harber}},
  \bibinfo{author}{\bibfnamefont{D.~L.} \bibnamefont{Whitaker}},
  \bibnamefont{and} \bibinfo{author}{\bibfnamefont{E.~A.}
  \bibnamefont{Cornell}}, \bibinfo{journal}{J. Low Temp. Phys.}
  \textbf{\bibinfo{volume}{132}}, \bibinfo{pages}{309} (\bibinfo{year}{2003}).

\bibitem[{\citenamefont{Leanhardt
  et~al.}(2003{\natexlab{b}})\citenamefont{Leanhardt, Shin, Chikkatur,
  Kielpinski, Ketterle, and Pritchard}}]{Leanhardt03}
\bibinfo{author}{\bibfnamefont{A.~E.} \bibnamefont{Leanhardt}},
  \bibinfo{author}{\bibfnamefont{Y.}~\bibnamefont{Shin}},
  \bibinfo{author}{\bibfnamefont{A.~P.} \bibnamefont{Chikkatur}},
  \bibinfo{author}{\bibfnamefont{D.}~\bibnamefont{Kielpinski}},
  \bibinfo{author}{\bibfnamefont{W.}~\bibnamefont{Ketterle}}, \bibnamefont{and}
  \bibinfo{author}{\bibfnamefont{D.~E.} \bibnamefont{Pritchard}},
  \bibinfo{journal}{Phys. Rev. Lett.} \textbf{\bibinfo{volume}{90}},
  \bibinfo{pages}{100404} (\bibinfo{year}{2003}{\natexlab{b}}).

\bibitem[{\citenamefont{Javanainen and Wilkens}(1997)}]{Javanainen97}
\bibinfo{author}{\bibfnamefont{J.}~\bibnamefont{Javanainen}} \bibnamefont{and}
  \bibinfo{author}{\bibfnamefont{M.}~\bibnamefont{Wilkens}},
  \bibinfo{journal}{Phys. Rev. Lett.} \textbf{\bibinfo{volume}{78}},
  \bibinfo{pages}{4675} (\bibinfo{year}{1997}).

\bibitem[{\citenamefont{Berman}(1994)}]{Berman94}
\bibinfo{editor}{\bibfnamefont{P.~R.} \bibnamefont{Berman}}, ed.,
  \emph{\bibinfo{title}{Cavity Quantum Electrodynamics}}
  (\bibinfo{publisher}{Academic Press}, \bibinfo{address}{San Diego},
  \bibinfo{year}{1994}).

\end{thebibliography}

\pagebreak

\begin{figure}[h]
\caption{Scale
drawing of the trap structure.  The main fields are generated by the
four horizontal rods, each of which is a coaxial pair.   A pair
consists of an outer conductor that is a 5-mm-diameter, 1-mm-wall
oxygen-free high-conductivity
copper tube, an alumina insulator, and an inner conductor that
is a 1.6-mm-diameter copper wire.  
The rods are held by two boron nitride blocks, which also support
the leads and circuit connections. 
The right block has been depicted
as transparent in order to display the arrangement of the conductors. 
The rod centers form a square 15
mm on a side and the blocks are spaced 5 cm apart. The function of
each of the conductors is described in Fig.~2.}
\end{figure}

\begin{figure}[h]
\caption{Current flow
through waveguide, indicated by thickened lines with
directional arrows. The four rods in Fig.~1 are depicted here as the
edges of a rectangular box. Circuits (a) and (b) refer to the
current through the inner conductors, which provides the oscillating
bias field. Circuit (c) is composed of the outer conductors, which
supply the confinement quadrupole field.  The end loops on circuit
(c) help to minimize the axial quadrupole field.}
\end{figure}

\begin{figure}[h]
\caption{Loading a
Bose-Einstein condensate into the waveguide. The sequence of
pictures show the trapped condensate as the static quadrupole field
is gradually turned off: (a) 29 G/cm (b) 19 G/cm (c) 9.7 G/cm (d)
3.9 G/cm (e) 1.9 G/cm (f) 0 G/cm. During the loading, the cloud
moves due to the centers of the external quadrupole and the
waveguide not being aligned. The bias field here is 3.69 G, and the
final trap frequencies are $\omega_x = 2\pi\times 11$~Hz, $\omega_z =
2\pi\times 6.2$~Hz, and $\omega_y = 2\pi\times 0.6$~Hz. Panel (g)
shows an atomic cloud after increasing the bias field to 20.5 G,
with trap frequencies $\omega_x = 2\pi\times6.0$~Hz, $\omega_y =
2\pi\times1.2$~Hz and $\omega_z = 2\pi\times3.3$~Hz. We estimate the
temperature of this cloud to be 850 pK.}
\end{figure}

\pagebreak

\epsfig{file=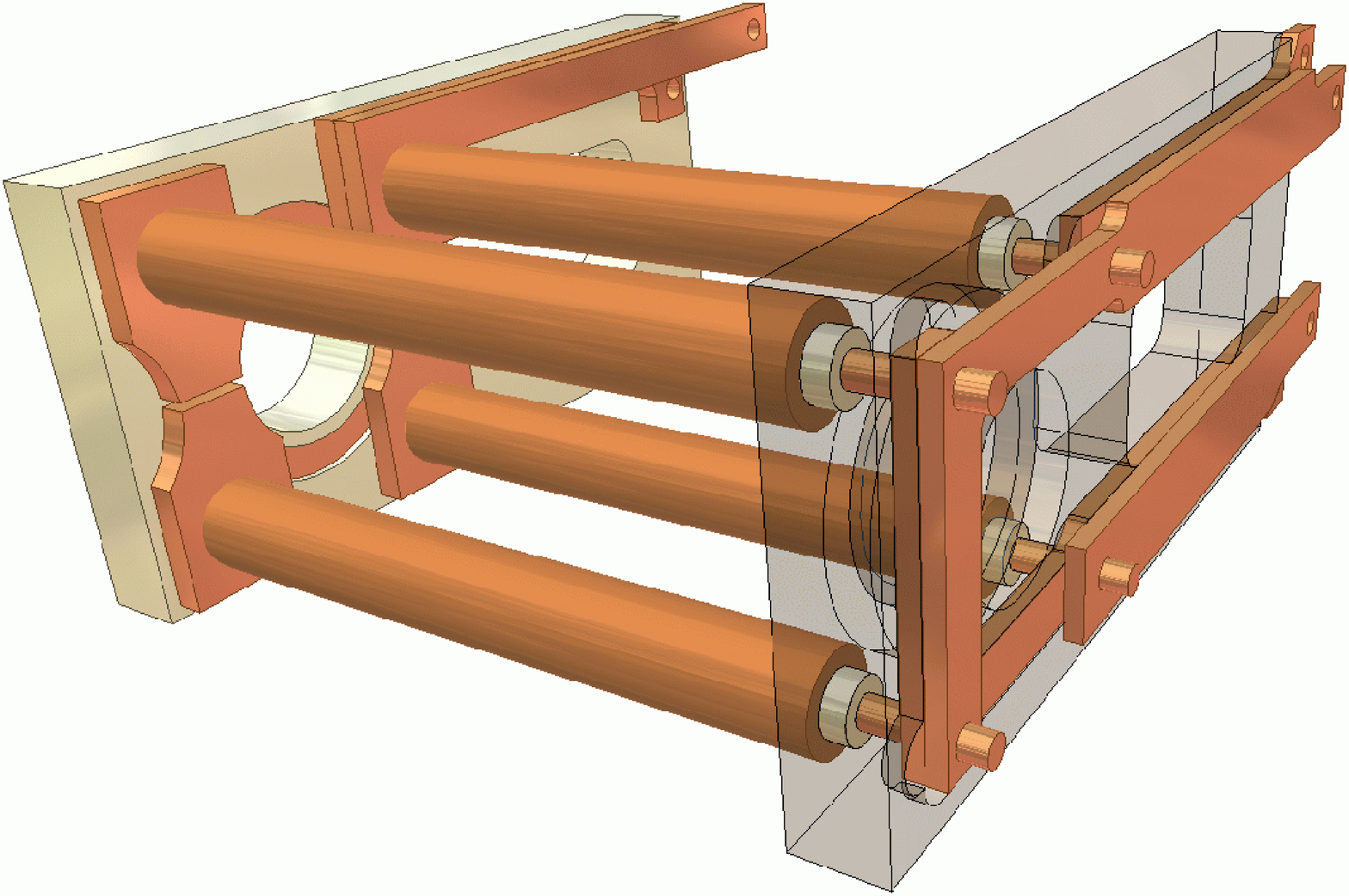,width=3.25in}

\pagebreak

\epsfig{file=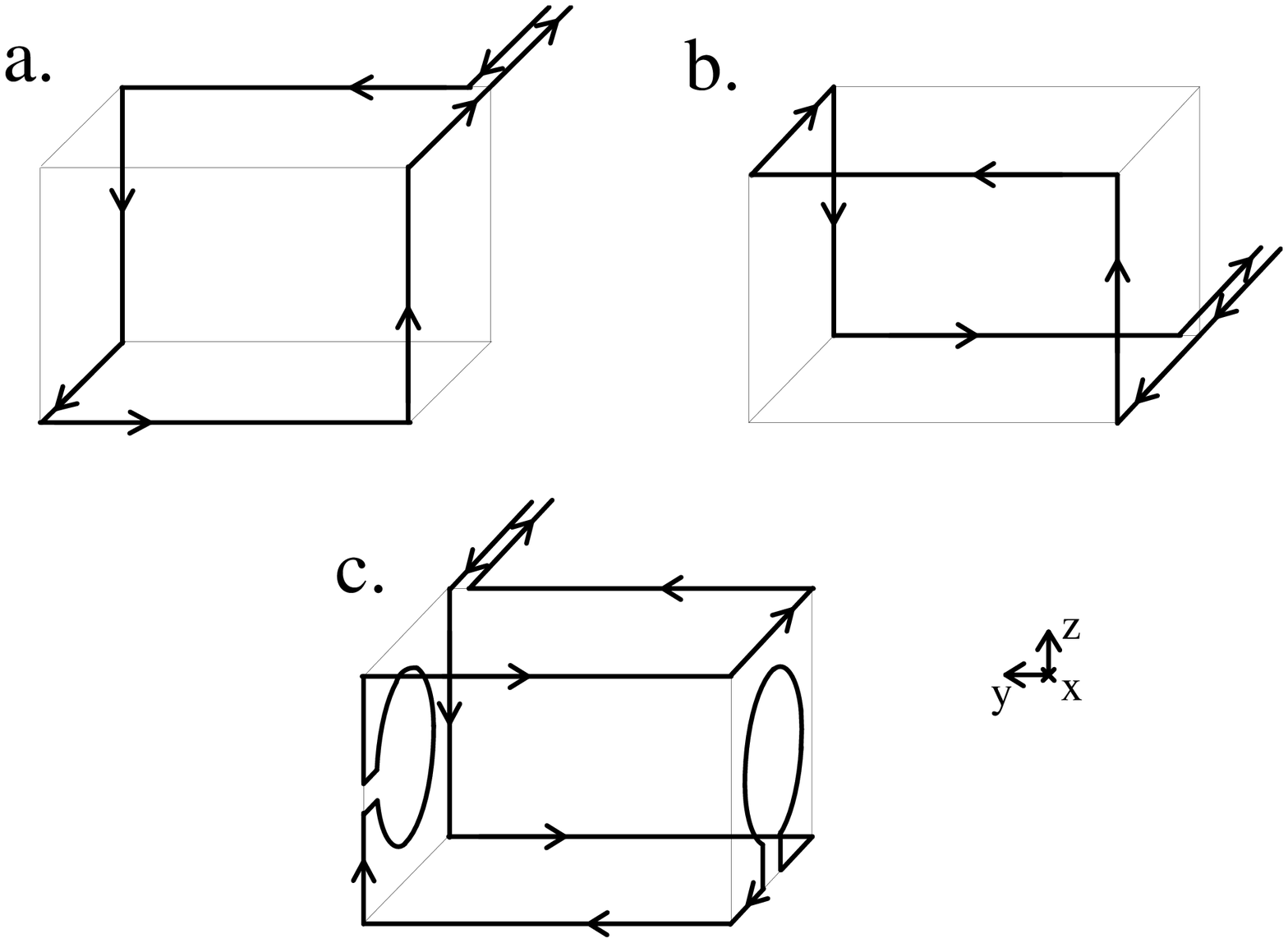,width=3.25in}

\pagebreak

\epsfig{file=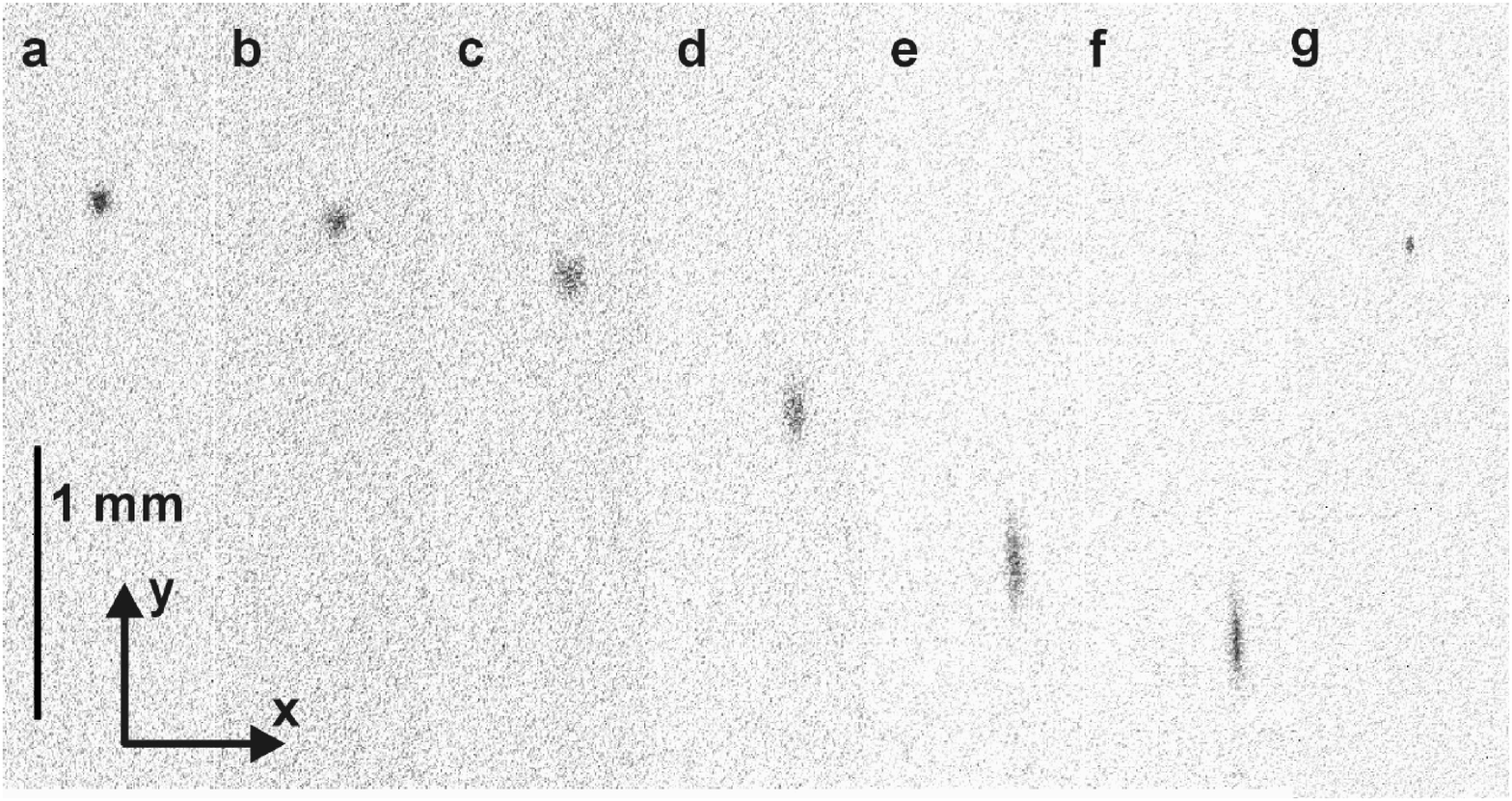,width=3.25in}

\end{document}